\documentclass[aps,prl,showpacs,twocolumn, superscriptaddress]{revtex4}

\usepackage{graphicx}
\usepackage{verbatim}

\begin{document}

\title{Observation of unconventional band topology in a superconducting doped topological insulator, Cu$_x$Bi$_2$Se$_3$: Topological or non-Abelian superconductor?}

\author{L. Wray}
\affiliation{Department of Physics, Joseph Henry Laboratories of
Physics, Princeton University, Princeton, NJ 08544, USA}
\affiliation{Advanced Light Source, Lawrence Berkeley National Laboratory, Berkeley, California 94305, USA}
\author{S. Xu}
\author{J. Xiong}
\author{Y. Xia}
\affiliation{Department of Physics, Joseph Henry Laboratories of
Physics, Princeton University, Princeton, NJ 08544, USA}
\author{D. Qian}
\affiliation{Department of Physics, Joseph Henry Laboratories of Physics, Princeton University, Princeton, NJ 08544, USA}
\affiliation{Department of Physics, Shanghai Jiao Tong University, Shanghai 200030, People's Republic of China}
\author{H. Lin}
\author{A. Bansil}
\affiliation{Department of Physics, Northeastern University, Boston, MA 02115, USA}
\author{Y. Hor}
\author{R. Cava}
\affiliation{Department of Chemistry, Princeton University, Princeton, NJ 08544, USA}
\author{M.Z. Hasan}
\affiliation{Department of Physics, Joseph Henry Laboratories of
Physics, Princeton University, Princeton, NJ 08544, USA}

\begin{abstract}

The Cu-doped topological insulator Bi$_2$Se$_3$ has recently been found to undergo a superconducting transition upon cooling, raising the possibilities that it is the first known ``topological superconductor" or realizes a novel non-Abelian superconducting state. Its true nature depends critically on the bulk and surface state band topology. We present the first photoemission spectroscopy results where by examining the band topology at many different copper doping values we discover that the topologically protected spin-helical surface states remain well protected and separate from bulk Dirac bands at the Fermi level where Copper pairing occurs in the optimally doped topological insulator. The addition of copper is found to result in nonlinear electron doping and strong renormalization of the topological surface states. These highly unusual observations strongly suggest that superconductivity on the topological surface of Cu$_x$Bi$_2$Se$_3$ cannot be of any conventional type in account of the general topological theory. Characteristics of the three dimensional bulk Dirac band structure are reported for the first time with respect to the superconducting doping state and topological invariant properties which should help formulate a specific theory for this novel superconductor.

\end{abstract}



\date{\today}

\maketitle


Topological insulators embody a new state of matter characterized by topological invariants of the band structure rather than spontaneously broken symmetry, and feature massless Dirac-like conduction states on their surfaces \cite{moore1, TIbasic,MooreAndBal,DavidNat1}. Bismuth selenide in particular has been found to be an ideal ``hydrogen atom" topological insulator, realizing the simplest known case of topologically nontrivial band structure \cite{MatthewNatPhys,DavidScience,DavidTunable,ZhangPred, BiTeSbTe,HorPtype,ZhangFilm}. It has been proposed that inducing a superconducting gap in the surface states of topological insulators will lead to fault tolerant non-Abelian surface physics with potential application in spintronics and quantum computing \cite{FuSCproximity}. In this Letter, we use angle resolved photoemission spectroscopy (ARPES) to examine the effect of copper doping on the electron dynamics of Cu$_x$Bi$_2$Se$_3$, which has been shown to lead to a bulk superconducting transition at x$\geq$0.1 (max T$_C$=3.8$^o$ K) \cite{HorSC}. Band structure in the normal state superconductor is found to preserve a minimalist scenario for a topological metal, consisting of a single bulk band with nearly isotropic massive Dirac-like dispersion and a well defined, topologically protected surface Dirac cone. As a result of this band structure, we find that superconductivity at the sample surface \emph{cannot be conventional}. Copper doping is systematically observed to have complex effects on the system, adding a small number of electrons and causing a strong renormalization of the surface bands in such a way as to preserve the topological character. The doped compound exhibits hexagonal dispersion anisotropy, with important implications for low energy interactions and potential device development \cite{FuHexagonal}.

\begin{figure}[t]
\includegraphics[width = 8.5cm]{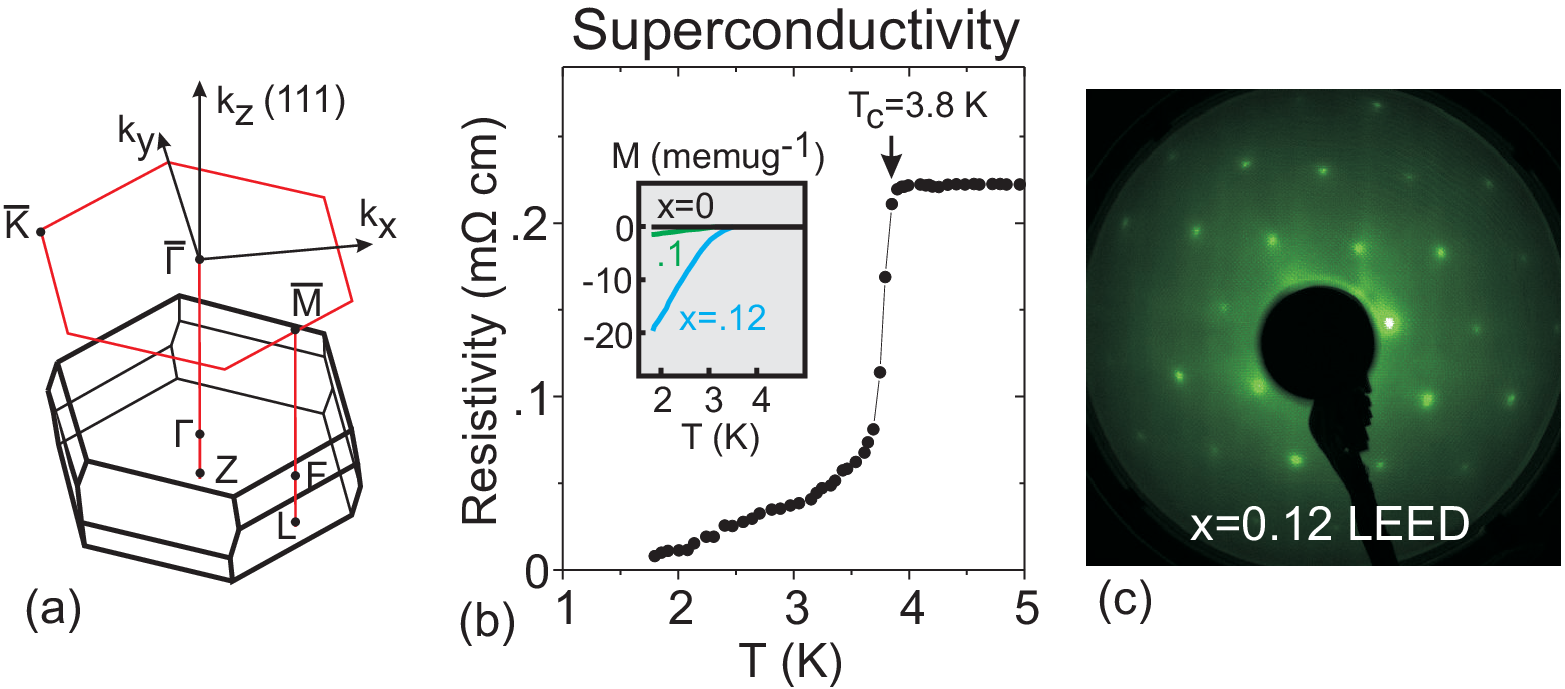}
\caption{{\bf{Superconductivity in a doped topological insulator Bi$_2$Se$_3$}}: (a) Resistivity and magnetic susceptibility measurements from Ref. \cite{HorSC} show a superconducting transition temperature of 3.8$^o$K at optimal copper doping (x=0.12). (b) The hexagonal surface state Brillouin zone of Cu$_x$Bi$_2$Se$_3$ is drawn in red above a diagram of the three dimensional bulk Brillouin zone.  (c) A LEED image taken at 200eV shows a well ordered surface with no signs of superstructure.}
\end{figure}

Undoped Bi$_2$Se$_3$ is a topological insulator with a large band gap ($>$300 meV) \cite{MatthewNatPhys}, and belongs to a class of materials M$_2$X$_3$ (M=Bi,Sb; X=S,Se,Te) that includes at least two other topologically nontrivial materials, Bi$_2$Te$_3$ and Sb$_2$Te$_3$, with smaller band gaps and more complicated band structure \cite{ZhangPred,BiTeSbTe}. These materials share a rhombohedral crystal structure, with a five atom unit cell arranged in quintuple layers, and have been investigated extensively in connection to thermoelectric applications \cite{DiSalvo}. Unlike band structure in topologically trivial materials, which is commonly more parabolic (``classical" E=$\frac{p^2}{2M}$), it is most natural for the bulk conduction bands of topological insulators to realize Dirac-like dispersion following an analogue of Einstein's equation for the energy of a relativistic particle (E$^2$=M$^2$v$_C^4$+p$^2$v$_C^2$) with a critical velocity ``v$_C$" analogous to the speed of light. The addition of electrons from copper doping allows us to quantitatively evaluate the bulk Dirac-like band character in the relativistic regime.

Angle resolved photoemission spectroscopy (ARPES) measurements were performed at the Advanced Light Source beamline 10.0.1 using 35.5-48 eV photons and Stanford Synchrotron Radiation Laboratory (7-22eV photons) with better than 15 meV energy resolution and overall angular resolution better than 1$\%$ of the Brillouin zone (BZ). Samples were cleaved and measured at 15$^o$K, in a vacuum maintained below 8$\times$10$^{-11}$ Torr. Momentum along the z-axis is determined using an inner potential of 9.5 eV, consistent with previous ARPES investigations of undoped Bi$_2$Se$_3$ \cite{MatthewNatPhys}.

\begin{figure*}[t]
\includegraphics[width = 12cm]{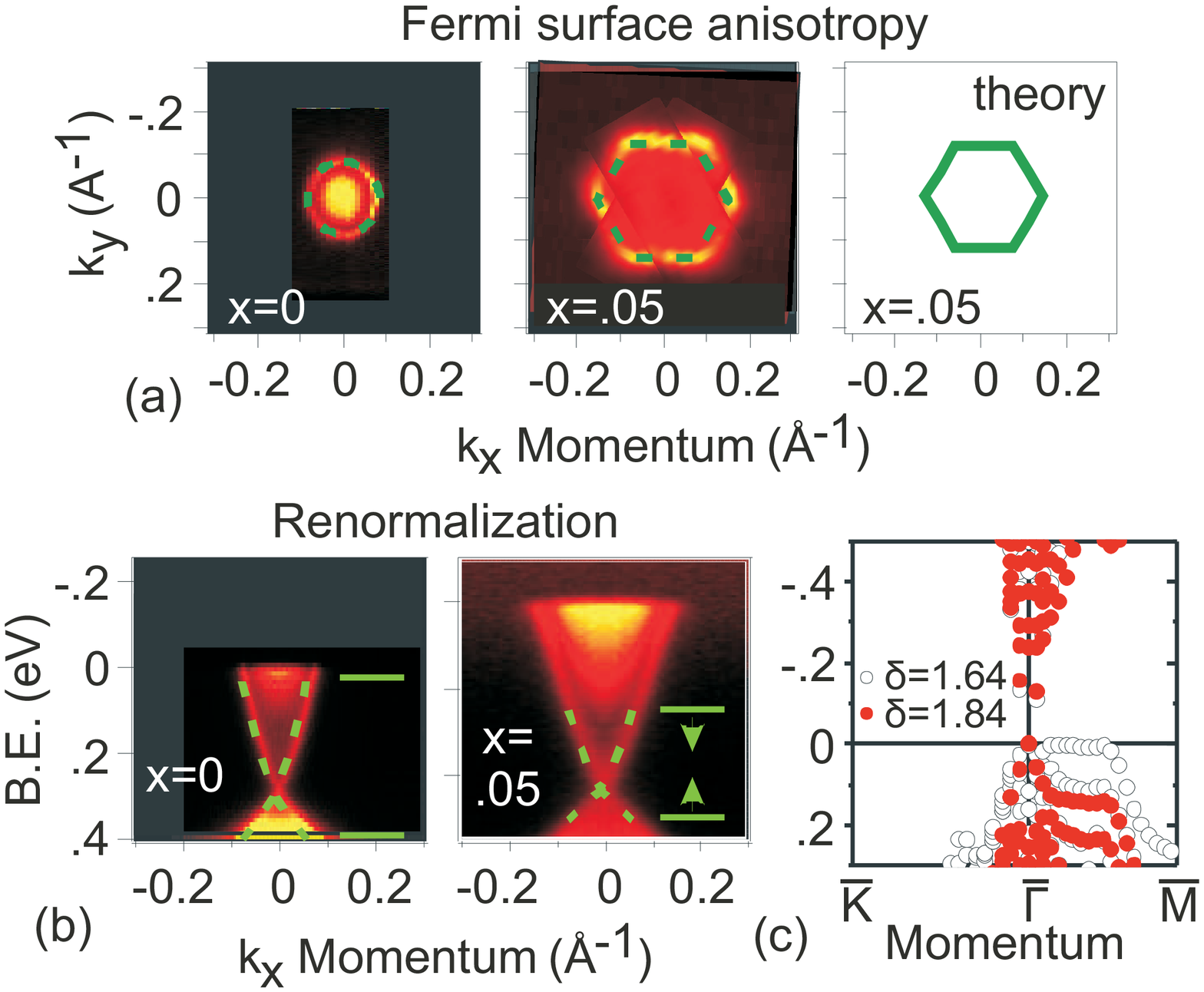}
\caption{{\bf{Nonlinear doping and band topology}}: (a) Symmetrized Fermi surfaces are displayed together with (right) an LDA prediction based on the Dirac point energy at x=0.05 copper doping. For clarity, the middle panel is measured at a photon energy that suppresses bulk resonance (48eV). (b) ARPES measurements are shown at high photon energy (E$>$20 eV) for non-superconducting Cu$_x$Bi$_2$Se$_3$. Particle velocities in the surface state upper Dirac cone are reduced by 30$\%$ after the addition of x=0.05 copper doping. (c) When the Bi-Se plane spacing at the surface of a 12-layer slab is increased by 0.2 $\AA$ (1.64 to 1.84 $\AA$), dispersion in the upper Dirac cone increases by 16$\%$, similar to the amount of renormalization we have observed in the superconducting compound. The surface state Dirac point is set at zero energy for each calculation. (d) The total number of charge carriers in the bulk and at the surface is calculated from the Luttinger count ($\frac{FS\,area}{BZ\,area}$, $\times$2 for the doubly degenerate bulk band).}
\end{figure*}

Surface and bulk state band calculations were performed for comparison with the experimental data, using the LAPW method implemented in the WIEN2K package \cite{wien2k}. Details of the calculation are identical to those described in Ref. \cite{MatthewNatPhys}.

To better understand the highly nonlinear doping effect of copper, we present data on several copper-added Cu$_x$Bi$_2$Se$_3$ crystals (x=0, 0.01, 0.05, 0.12) and copper substituted Cu$_{0.1}$Bi$_{1.9}$Se$_3$, grown as described in Ref. \cite{HorSC}. Copper atoms intercalated between the van der Waals bonded selenium planes are thought to be single electron donors, while substitutional defects in which copper replaces bismuth in the lattice (Cu$_{Bi}$) are expected to each contribute two holes to the system \cite{CuAmphoteric}. Introducing x=0.12 copper doping for optimal superconductivity has been found to shift the z-axis lattice parameter by only 1.5$\%$ while leaving the in-plane lattice parameters and long-range crystalline order intact \cite{HorSC}.

\begin{figure*}[t]
\includegraphics[width = 13cm]{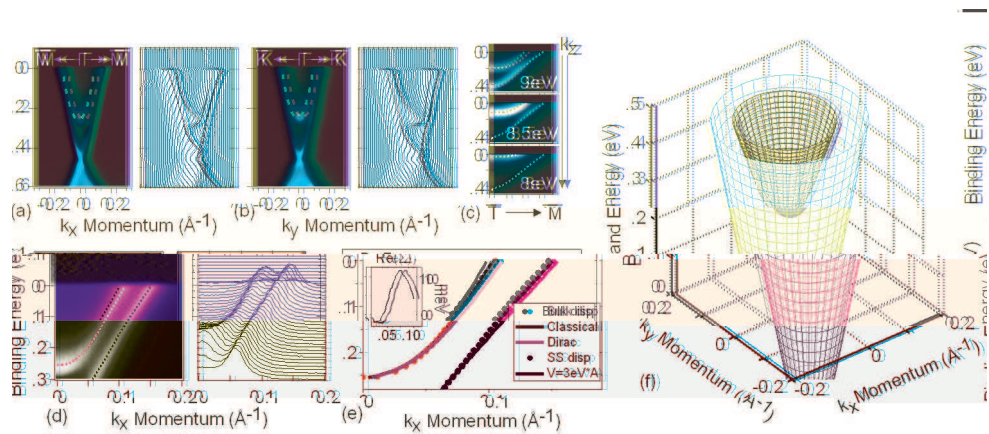}
\caption{{\bf{Bulk Dirac states vs Surface band topology and superconductivity}}: (a-b) Momentum dependence of the bulk and surface conduction bands in superconducting Cu$_{0.12}$Bi$_2$Se$_3$ is measured through the 3D Brillouin zone center with low energy (9.75eV) photons for enhanced bulk sensitivity. (c) Small panels show that the bulk and surface bands remain separate at intermediate k$_z$ values. (d) Dispersion is traced on a close-up image of the conduction bands. (e) Energy-momentum dispersion of the bulk electrons is compared with Dirac-like (v$_c$=6 eV$\cdot$$\AA$) and classical (parabolic) fits with an effective mass of 0.155m$_e$. Neither fit can perfectly account for a slight bend in dispersion near 90 meV binding energy. An inset shows the difference between bulk band energy and the Dirac-like fit curve. (f,blue) The bulk band dispersion approximated from ARPES data is plotted at k$_z$=0, surrounded by (green) a massless ``light-like" limit, illustrating that bulk electron kinetics are in the ``relativistic" Dirac regime.}
\end{figure*}

The qualitative effect of copper addition is an enlargement of the Fermi surface from electron doping and, surprisingly, a strong renormalization of the surface state. The surface state dispersion is flattened by 30$\%$ when copper doping of x=0.05 is added to the stoichiometric compound, and the Fermi surface becomes hexagonally anisotropic in a way consistent with numerical predictions in the local density approximation (LDA). Tunability of the surface state anisotropy is important for control of unconventional ordered states that may appear uniquely in topologically ordered materials \cite{FuHexagonal}.

The bulk and surface-derived band structure of the normal state optimally doped superconductor (x=0.12) are explored in Fig-3. The bottom of the conduction band for electrons in the bulk is found at the three dimensional $\Gamma$-point (k$_z$=k$_x$=k$_y$=0), and can be seen inside the upper surface state (SS) cone. Due to the renormalization effect and less-than-linear electron doping, the occupied surface state band structure does not intersect with the bulk conduction band structure at any point in momentum space, making the surface state ``well defined" and topologically protected. The bulk conduction band is only clearly visible when low photon energies (h$\nu$$<$20eV) are used to increase bulk penetration, probably due to a screening effect related to excess negative charge carriers in the surface state. (see analysis in Fig-2(d)) The gap between bulk valence and conduction bands appears to be unchanged upon copper doping.

Using 9.75eV photons to view the $\Gamma$-$\overline{M}$ and $\Gamma$-$\overline{K}$ directions shows Fermi momenta of 0.110$\pm$3 $\AA^{-1}$ and 0.106$\pm$3 $\AA^{-1}$ respectively. Varying incident energy to observe dispersion along the $\hat{z}$ axis ($\Gamma$-Z direction) reveals a Fermi momentum of 0.12 $\AA^{-1}$, suggesting that the bulk electron kinetics are three dimensionally isotropic. Carefully tracing the band (Fig-3(d-e)) yields a Fermi velocity of 3.5 eV$\times\AA$ along $\Gamma$-$\overline{M}$ and 4.1 eV$\times\AA$ along $\Gamma$-$\overline{K}$, estimated within 50 meV of the Fermi level. Assuming that the bulk conduction band forms a Fermi sea for superconductivity, the superconducting correlation length can be estimated based on the average Fermi velocity and superconducting critical temperature to be about $2000\AA$ ($\xi_0$$\sim0.2\times\hbar v_F/K_BT_C=0.2\times3.8eV\AA/(K_B\times3.8^oK)=2000\AA$), large enough for phase fluctuations to be neglected in the neighborhood of T$_C$. This is a typical coherence length for conventional superconducting materials, and much greater than that seen in other unconventional superconductors, such as strongly correlated cuprates ($\xi_0$$\sim100-200\AA$) or cobaltates ($\xi_0$$\sim200\AA$) \cite{SCproperties}.

Along both the $\Gamma$-$\overline{M}$ and $\Gamma$-$\overline{K}$ directions, dispersion of the bulk conduction band appears to approach a limiting velocity at large momentum, rather than following a parabolic arc. This behavior, like the ``massless" linear dispersion of the surface state bands, is mathematically analogous to the energy-momentum relationship of a relativistic particle approaching the speed of light. Classical (paraboloic, m=0.155 m$_e$) and relativistic Dirac-like (m=0.155 m$_e$, v$_c$=6 eV$\cdot\AA$) energy dispersions are plotted as fits for the bulk band in Fig-3(e). We find that a relativistic fit can more accurately reproduce the observed dispersion, however neither fit can account for a slight bend in the dispersion centered near 90 meV, which may suggest strong electron-phonon interactions in the system consistent with phonon-mediated superconductivity. The simplicity of band structure in the doped Bi$_2$Se$_3$ system clearly sets it apart from other known topological insulators (e.g. Bi$_2$Te$_3$) that have far greater deviation from pure Dirac-like kinetics \cite{DavidTunable,BiTeSbTe,FuHexagonal,ChenBiTe}.

The weakness of copper doping is critical to preserving topological order in the superconducting state, by keeping the Fermi level beneath the point of intersection between bulk and surface conduction bands. (near the top of the diagram in Fig-4(a,right)) The number of conducting charge carriers at the surface and in the bulk are estimated using the Luttinger count in Fig-2(d), by dividing the surface and bulk Fermi surface areas by the total area of the Brillouin zone. Although the doping increases monotonically as copper is added, the carrier density in the bulk is only $\sim$1/30th of what would be expected if all copper were intercalated between paired selenium layers. It is therefor likely that nearly one third of the added copper enters the sample through substitutional defects with bismuth, adding holes that counterbalance most of the electron doping from intercalation.

\begin{figure}[t]
\includegraphics[width = 8.7cm]{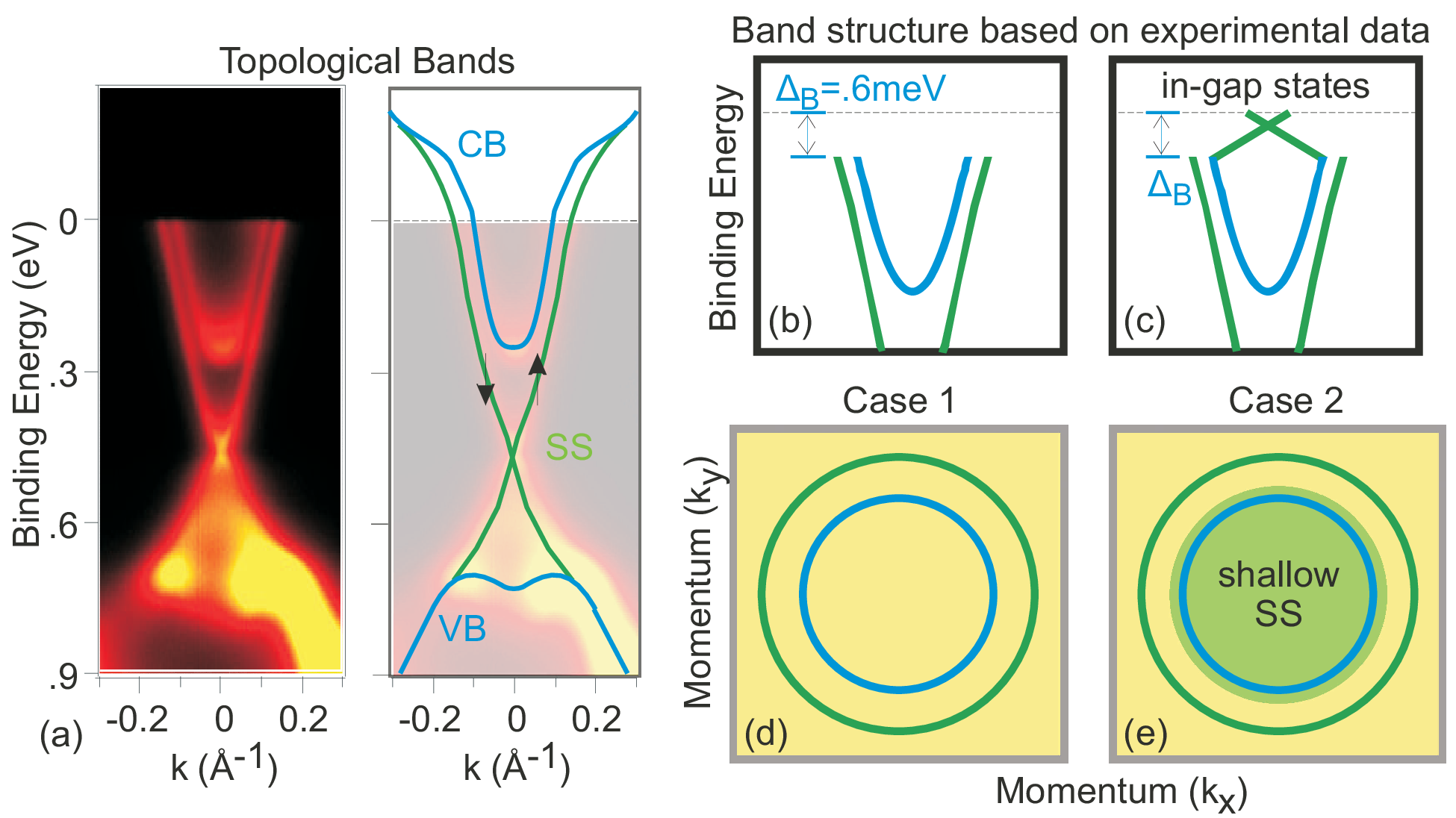}
\caption{{\bf{Symmetry breaking and the topological surface:}} (a) Topologically protected surface states cross the Fermi level before merging with the bulk valence and conduction bands in a topological material. (b) If superconducting parity is even, the surface states will be gapped due to their proximity to the bulk superconducting wavefunction, allowing non-Abelian Majoran Fermion surface vortices. (c) If parity is odd the material will be a topological superconductor, and new states will appear below T$_C$ to span the bulk superconducting gap. (d,e) States within 5meV of the Fermi level are shaded for the k$_z$=0 plane of (Case 1) a fully gapped doped topological insulator and (Case 2) one example of a topological superconductor.}
\end{figure}

Attempting to force the creation of Cu$_{Bi}$ defects by adding less bismuth results in very weak hole doping for Cu$_{0.1}$Bi$_{1.9}$Se$_3$. In this case, the binding energy of the Dirac point decreases by more than 100 meV relative to the undoped compound, raising the bulk conduction band entirely above the Fermi level so that the material is a traditional topological insulator. An explanation for why hole doping is not stronger in this case may be that defects in which selenium fills bismuth vacancies will result in the addition of electrons.

The surface state renormalization that we have observed is also instrumental in separating the bulk and surface band structure. Renormalization is likely to result from a combination of increased carrier density on the surface relative to the bulk, evident from Fig-2(d), and relaxation of interlayer bond lengths near the cleaved surface. Changes to interatomic bonding lengths at the surface of strongly spin-orbit coupled systems can have a significant effect on surface state dispersion \cite{BiDistortion}, and could be effected by the surface carrier density and presence of copper. We find that increasing the distance between the outermost bismuth and selenium layers by 0.2 $\AA$ can account for the difference in dispersion between undoped and superconducting doped crystals (Fig-2(c)).

A cartoon of band structure in a topological metal (insulator) is overlaid on ARPES data in Fig-4(a), illustrating that the gapped bulk bands are connected by a singly degenerate gapless surface state. When a bulk superconducting transition is introduced, there are two likely scenarios. If the parity eigenvalue of the bulk superconducting state is even (Fig-4(b,d)), electrons in the surface state will participate in bulk superconductivity through the proximity effect, resulting in the appearance of non-Abelian vortex states of great interest for quantum computing \cite{FuSCproximity}. If parity is odd, the system will be a topological superconductor \cite{FuNew}, with new gapless states appearing beneath the T$_C$. ARPES measurements cannot resolve the bulk superconducting gap, which is expected to be only $\sim$0.6 meV from BCS theory (3.5$\times$K$_B$T$_C$/2=0.6 meV).

Although our results show that Bi$_2$Se$_3$ is a nearly ideal minimalist topological insulating system, many of the most interesting physical properties emerge due to its deviations from the simplest case of perfectly isotropic Dirac-like electron kinetics. Recent theoretical explorations have suggested that magnetic perturbations will not readily open a mass gap in the surface state of a topological insulator with perfectly linear dispersion \cite{ImpurityStates,TopoFieldTheory}, but can do so when the bands have some upward concavity as in Bi$_2$Se$_3$ \cite{FerroPosMassBiSe}. The hexagonal Fermi surface anisotropy demonstrated upon copper doping adds an out of plane component to spin polarization, and is a likely precursor to novel two dimensional ordered states unique to the surfaces of materials with topological order \cite{FuHexagonal}. Furthermore, our observation of large surface charge contained within the topologically protected surface state and a dramatic band renormalization effect have important implications for manipulation of topological insulator systems in future experiments and for technological applications.

In summary, we have investigated the surface and bulk band topology of the superconducting doped topological insulator Cu$_x$Bi$_2$Se$_3$. We discover that the copper-doped superconducting state unexpectedly preserves the band structure and topological order of undoped bismuth selenide, and that superconductivity at the sample surface cannot be conventional. The bulk conduction band conforms approximately to a massive Dirac dispersion, and has a slight bend near 90 meV that may be indicative of electron-boson coupling. We also observe strong doping dependent renormalization of surface states and hexagonal anisotropy that preserve the topological order via a $\pi$ Berry's phase invariant and are of interest with respect to spin transport properties and potential device development.

\begin{acknowledgments}

We thank helpful conversation with L. Fu, B. A. Bernevig and P.W. Anderson.

\end{acknowledgments}

\begin{figure}[t]
\includegraphics[width = 8.5cm]{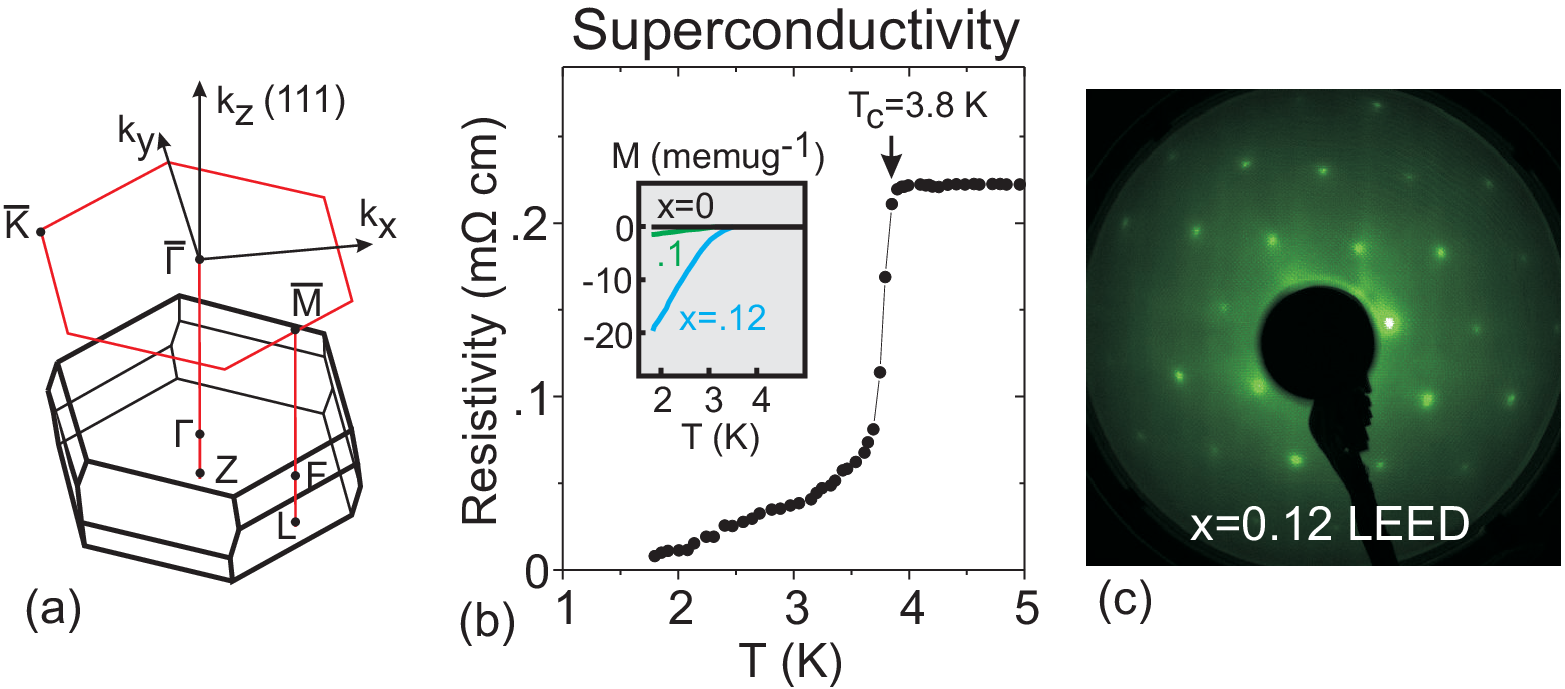}
\caption{{\bf{Superconductivity in a doped topological insulator Bi$_2$Se$_3$}}: (a) Resistivity and magnetic susceptibility measurements from Ref. \cite{HorSC} show a superconducting transition temperature of 3.8$^o$K at optimal copper doping (x=0.12). (b) The hexagonal surface state Brillouin zone of Cu$_x$Bi$_2$Se$_3$ is drawn in red above a diagram of the three dimensional bulk Brillouin zone.  (c) A LEED image taken at 200eV shows a well ordered surface with no signs of superstructure.}
\end{figure}

\begin{figure}[t]
\includegraphics[width = 12cm]{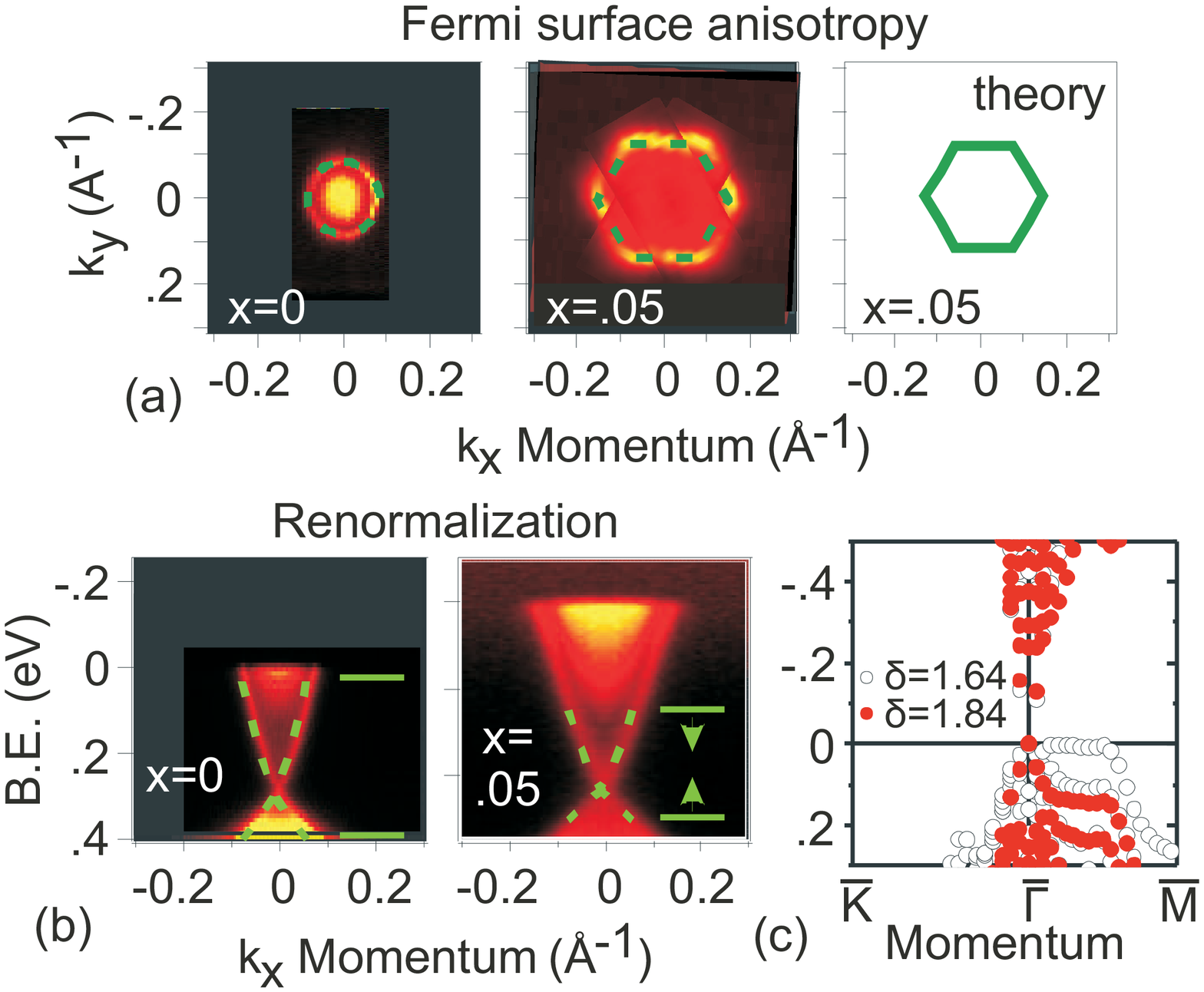}
\caption{{\bf{Nonlinear doping and band topology}}: (a) Symmetrized Fermi surfaces are displayed together with (right) an LDA prediction based on the Dirac point energy at x=0.05 copper doping. For clarity, the middle panel is measured at a photon energy that suppresses bulk resonance (48eV). (b) ARPES measurements are shown at high photon energy (E$>$20 eV) for non-superconducting Cu$_x$Bi$_2$Se$_3$. Particle velocities in the surface state upper Dirac cone are reduced by 30$\%$ after the addition of x=0.05 copper doping. (c) When the Bi-Se plane spacing at the surface of a 12-layer slab is increased by 0.2 $\AA$ (1.64 to 1.84 $\AA$), dispersion in the upper Dirac cone increases by 16$\%$, similar to the amount of renormalization we have observed in the superconducting compound. The surface state Dirac point is set at zero energy for each calculation. (d) The total number of charge carriers in the bulk and at the surface is calculated from the Luttinger count ($\frac{FS\,area}{BZ\,area}$, $\times$2 for the doubly degenerate bulk band).}
\end{figure}

\begin{figure}[t]
\includegraphics[width = 13cm]{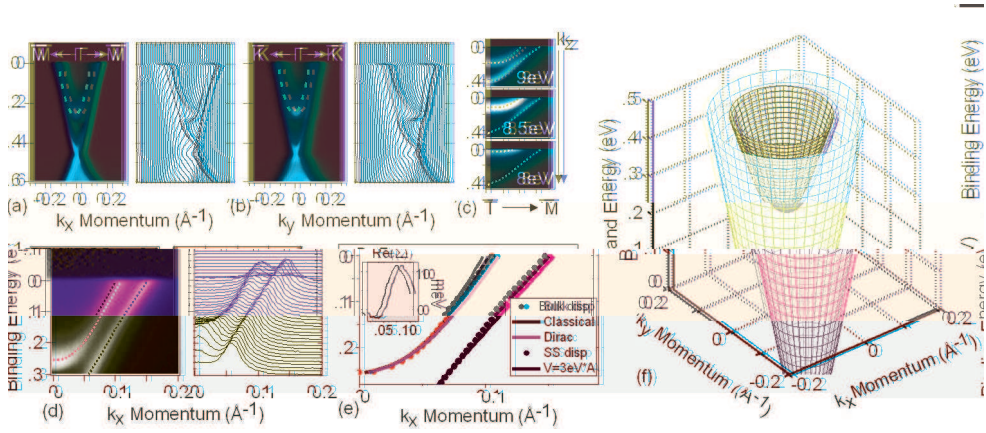}
\caption{{\bf{Bulk Dirac structure vs Surface band topology and superconductivity}}: (a-b) Momentum dependence of the bulk and surface conduction bands in superconducting Cu$_{0.12}$Bi$_2$Se$_3$ is measured through the 3D Brillouin zone center with low energy (9.75eV) photons for enhanced bulk sensitivity. (c) Small panels show that the bulk and surface bands remain separate at intermediate k$_z$ values. (d) Dispersion is traced on a close-up image of the conduction bands. (e) Energy-momentum dispersion of the bulk electrons is compared with Dirac-like (v$_c$=6 eV$\cdot$$\AA$) and classical (parabolic) fits with an effective mass of 0.155m$_e$. Neither fit can perfectly account for a slight bend in dispersion near 90 meV binding energy. An inset shows the difference between bulk band energy and the Dirac-like fit curve. (f,blue) The bulk band dispersion approximated from ARPES data is plotted at k$_z$=0, surrounded by (green) a massless ``light-like" limit, illustrating that bulk electron kinetics are in the ``relativistic" Dirac regime.}
\end{figure}

\begin{figure}[t]
\includegraphics[width = 8.7cm]{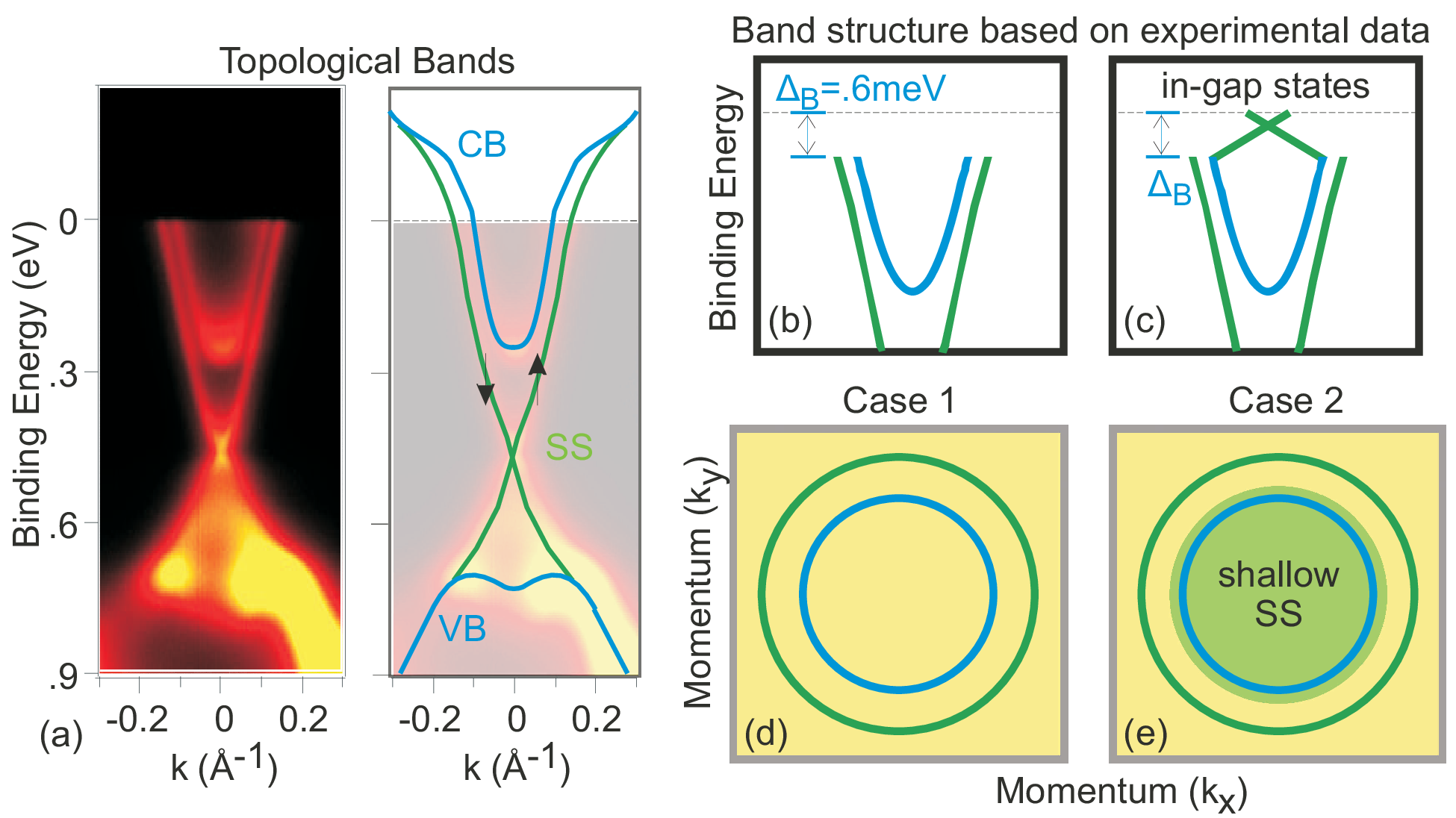}
\caption{{\bf{Symmetry breaking and the topological surface:}} (a) Topologically protected surface states cross the Fermi level before merging with the bulk valence and conduction bands in a topological material. (b) If superconducting parity is even, the surface states will be gapped due to their proximity to the bulk superconducting wavefunction, allowing non-Abelian Majorana Fermion surface vortices. (c) If parity is odd the material will be a topological superconductor, and new states will appear below T$_C$ to span the bulk superconducting gap. (d,e) States within 5meV of the Fermi level are shaded for the k$_z$=0 plane of (Case 1) a fully gapped doped topological insulator and (Case 2) one example of a topological superconductor.}
\end{figure}

\end{document}